# Antioxidant capacity is repeatable across years but does not consistently correlate with a marker of peroxidation in a free-living passerine bird.

Running title: Correlation of oxidative state markers

Charlotte Récapet*[1,2,3], Mathilde Arrivé[4], Blandine Doligez[2,5†] and Pierre Bize[6†]

[1] UMR 1224 ECOBIOP; Université de Pau & Pays Adour, INRA; Saint-Pée-sur-Nivelle, France.

[2] UMR 5558 LBBE; Université de Lyon, Université Lyon 1, CNRS; Villeurbanne, France.

[3] Département d'Ecologie et d'Evolution, Université de Lausanne, Switzerland.

[4] UPR 2357 IBMP ; Université de Strasbourg, CNRS ; Strasbourg, France.

[5] Animal Ecology, Department of Ecology and Genetics, Evolutionary Biology Centre, Uppsala University, Uppsala, Sweden.

[6] School of Biological Sciences, Zoology Building, University of Aberdeen, Aberdeen, United-Kingdom.

† These authors contributed equally to the work.

* Corresponding author: charlotte.recapet@univ-pau.fr, phone +33 559574266 or +33 559515978.

ORCID for Charlotte Récapet (0000-0001-5414-8412), Blandine Doligez (0000-0003-3015-5022) and Pierre Bize (0000-0002-6759-4371).

**Authors' contributions**

C.R., B.D. and P.B. designed the study. C.R. carried out the fieldwork. C.R. and M.A. performed the laboratory analyses and analysed the data. C.R., M.A., B.D. and P.B. drafted the manuscript. All authors gave final approval for publication.

**Acknowledgments**

Many thanks to the landowners of Gotland for access to the study sites; to Lars Gustafsson for logistics on the field; to fifteen students and field assistants for their contribution to fieldwork; to François Criscuolo for his help designing the study and his comments on this manuscript; and to Alan A. Cohen and an anonymous reviewer for their comments on a previous version of this manuscript.

This work was supported by the French National Center for Scientific Research (PICS France-Switzerland to B.D.); the French Ministry of Research (PhD fellowship to C.R.); the University of Aberdeen (stipend to C.R.); the L'Oréal Foundation-UNESCO "For Women in Science" program (fellowship to C.R.); the Région Rhône-Alpes (Explora'doc mobility grant to C.R.); the Fédération de Recherche 41 BioEnvironnement et Santé (training grant to CR); the Rectors' Conference of the Swiss Universities (joint doctoral program grant to C.R.) and the Fondation pour l'Université de Lausanne (exchange grant to C.R.); and the Journal of Experimental Biology (travel grant to CR).

# Abstract

Oxidative stress occurs when reactive oxygen species (ROS) exceed antioxidant defences and have deleterious effects on cell function, health and survival. Therefore, organisms are expected to finely regulate pro-oxidant and antioxidant processes. ROS are mainly produced through aerobic metabolism and vary in response to changes in energetic requirements, whereas antioxidants may be enhanced, depleted or show no changes in response to changes in ROS levels. We investigated the repeatability, within-individual variation and correlation across different environmental conditions of two plasmatic markers of the oxidative balance in free-living adult collared flycatchers (*Ficedula hypoleuca*). We manipulated energetic constraints through increased flight costs in 2012 and 2013 and through a food supplementation in 2014. We then tested the relative importance of within- and between-individual variation on reactive oxygen metabolites (ROMs), a marker of lipid and protein peroxidation, and on non-enzymatic antioxidant defences (OXY test). We also investigated whether the experimental treatments modified the correlation between markers. Antioxidant defences, but not ROMs, were repeatable within and between years. Antioxidants increased under the food supplementation treatment but did not vary between breeding stages. ROMs increased during reproduction in females and were higher in females than males. Antioxidant defences and ROM concentration were globally positively correlated, but the correlation structure varied between experimental conditions and between years. Understanding the role of oxidative balance in wild animals will thus require a flexible, mechanistic approach to modelling their interactions.

# Introduction

Reactive oxygen species (ROS) produced during aerobic respiration are important actors of cell signalling pathways and immune responses (Metcalfe and Alonso-Alvarez 2010; Sena and Chandel 2012), but can have deleterious effects by oxidising macromolecules and thereby disrupting cell function (Avery 2014). Oxidative stress, resulting from an excess of ROS relative to antioxidant defences, is thus proposed as a credible mechanism underlying life-history trade-offs (Monaghan et al. 2009; Metcalfe and Alonso-Alvarez 2010). To understand the potential role of oxidative stress as a constraint on life-history traits, it is crucial to understand how organisms co-regulate pro- and anti-oxidant molecules. There is no *a priori* expectation on how measures of antioxidant defences should relate to measures of ROS levels and oxidative damages (Costantini and Verhulst 2009). The antioxidant capacity could be modulated to counteract the deleterious effects of ROS, thus resulting in a redox balance (Blount et al. 2016). Such compensation by an increase in antioxidant in response to ROS production would lead to a positive, or an absence of, correlation. Conversely, a deficit of energetic resources or dietary antioxidants might constrain antioxidant defences and thus decrease their ability to counteract ROS effects. If circulating antioxidants are then used up to protect the organism against increased reactive oxygen species, they should covary negatively with ROS production or oxidative damages. However, it is not clear whether antioxidant protection, especially through antioxidant enzymes, is energetically costly (Speakman and Krol 2010; Isaksson et al. 2011; Gems and Partridge 2013) and measured antioxidant defences could increase with ROS production, as more antioxidant are mobilized.

Variation in ROS production is also insufficiently understood. ROS are often expected to increase in response to increased metabolism, but the generality of this relationship is far from clear (Speakman and Selman 2011). A higher metabolic and respiratory rate might not be associated with a higher ROS production (Barja 2007; Glazier 2015; Salin et al. 2015b). First, the natural or experimental inhibition of mitochondrial respiration simultaneously slows down the electron flow through the electron transport chain, which is then in a reduced state, and increases the intra-mitochondrial concentration of oxygen, two mechanisms which promote ROS production (Bonawitz et al. 2007; Salin et al. 2015a). Second, uncoupling proteins (UCPs) and changes in the mitochondrial inner membrane structure modulate the inner membrane conductance to protons and in turn the relationship between energy consumption and ROS production (Brand 2000; Criscuolo et al. 2005; Stier et al. 2014a).

The interpretation of oxidative stress measures in ecological studies therefore requires a better understanding of their variability in the wild and the actual co-variation between markers. In particular, relationships between traits can differ greatly when measured at the within-individual level, where life-history trade-offs can be expressed, or at the between-individual level, where differences in individual quality or permanent environment are expected to play a large role (Stearns 1989; Wilson and Nussey 2009). To explore these questions, we measured two plasmatic markers of oxidative state, namely (i) organic hydroperoxides, acting as precursors of long-term oxidative damage, through the d-ROM test and (ii) non-enzymatic antioxidant capacity through the OXY test, in adults of a small passerine bird, the collared flycatcher *Ficedula albicollis* (Temminck, 1815), during three consecutive breeding seasons. These two markers, frequently used in wild bird populations, are sensitive to manipulations of energy expenditure or mitochondrial ROS production (Stier et al. 2014b; Vaugoyeau et al. 2015; Récapet et al. 2016c, 2017). They have been linked to dietary antioxidant intake (Beaulieu and Schaefer 2014), to reproductive effort (Beaulieu et al. 2011; Markó et al. 2011; Reichert et al. 2014; Wegmann et al. 2015) and to fitness outcomes (Geiger et al. 2012; Herborn et al. 2016). Although markers of oxidative state are often variable across tissues (Veskoukis et al. 2009), we focussed on plasma here because (i) it can be repeatedly sampled within and between years and (ii) we can expect quick variation of oxidative balance in the blood in response to energetic constraints (Nikolaidis et al. 2008). We first estimated the repeatability of oxidative state markers within the same season, as well as between seasons. Then we assessed the correlation between these markers while controlling for differences between years or between sexes that can spuriously generate strong correlations between markers (Christensen et al. 2015). If non-enzymatic antioxidants are constrained by availability in the diet, we expect a negative correlation between antioxidant capacity and reactive oxygen metabolites and this negative correlation should be stronger when feeding conditions are poorer. Conversely, if non-enzymatic antioxidants are up regulated at low costs in response to increased ROS production, we expect a positive correlation between these two markers, independently of the feeding conditions.

To experimentally test the effect of energetic and nutritional constraints on oxidative state, we manipulated wing load in breeding females (i.e. handicapped females) to increase their energy expenditure in 2012 and 2013, and we food supplemented breeding pairs during the nestling feeding stage to alleviate nutritional constraints and foraging energy expenditure in 2014. If increased energy expenditure translates into increased ROS production, we expect higher reactive oxygen metabolites when energetic constraints are stronger, i.e. (i) for handicapped females (compared to control ones) and (ii) for unfed pairs (compared to food supplemented ones). More generally, we also expect the correlations between the two markers of oxidative state to be stronger in handicapped females compared to control ones and in unfed pairs compared to food supplemented ones.

## **Material and methods**

*Study population*

The study was conducted during spring 2012 to 2014 in a natural population of collared flycatchers breeding on the island of Gotland, Sweden (57°07′N, 18°20′E). This hole-nesting bird readily breeds in the artificial nest boxes erected in the nine forest plots used for the study (between 13 and 78 nest boxes per plot distributed homogeneously in space). Nests were visited regularly to estimate laying date, clutch size and hatching date. Females were first caught during incubation (on average ± S.D. 7.2 ± 1.2 days after the start of incubation), before the start of any experimental treatment, then males and females were both caught when feeding nestlings (on average when nestlings were 9.0 ± 2.1 days old). For females, the interval between both captures was thus 13.6 ± 2.5 days on average (± S.D.). Upon capture, birds were ringed if previously unringed and blood sampled (see below).

*Experimental manipulations of energetic constraints*

In 2012 and 2013, we increased energetic constraints on females from the second half of the incubation period by cutting the two innermost primaries of each wing at their base, to mimic feather loss at the onset of moult (Moreno et al. 1999; Sanz et al. 2000; Ardia and Clotfelter 2007; Récapet et al. 2016c). Upon capture during incubation, females were assigned to the manipulated (handicapped) or the control group (same handling conditions but no feathers cut) alternatively by blocks of two females to avoid the two experimental groups to differ in treatment date. The manipulation was successful at increasing female energy expenditure, measured through the doubly-labelled water method (Récapet et al. 2016c). We did not manipulate the wing load of males in this experiment and found no effect of our manipulation on males, for example through compensatory behaviour (Récapet et al. 2016c).

Conversely, in 2014, we relieved energetic constraints on both parents during the nestling feeding period by providing additional food (Récapet et al. 2016a, 2017). When nestlings were two days old, transparent plastic containers were attached to the front side of the nest box. For supplemented pairs, 30g live maggots (larvae of *Calliphora erythrocephala*, Fibe AB, Kungsängen, Sweden) were placed in the containers once a day until nestlings were 12 days old (i.e. over a total of 11 days). This corresponded to approx. 150 individual larvae of 200mg per day, i.e. 25 larvae per nestling for a brood of 6 nestlings. Thus, the food supplementation spared about half of the approx. 360 daily parental visits estimated in control nests in the same year (Récapet et al. 2017). Control pairs received no food, but were visited daily to control for human disturbance. Pairs were assigned either to the control or supplemented group alternatively in space for a given hatching date, so as to distribute treatments homogenously in space both within and between study plots and in time within the breeding season.

*Measures of markers of the oxidative balance*

To measure blood markers of oxidative state, a 40μL blood sample was taken from the brachial vein into heparin-coated Microvettes (Sarstedt, Nümbrecht, Germany). Blood samples were maintained at 5°C in the field before being centrifuged in the evening to separate plasma from red blood cells. Plasma and red blood cells were then stored at -80°C until being analysed in the laboratory. A total of 860 blood samples was collected on nestling feeding males and females, and 256 on incubating females. Because of the low amount of plasma available for each sample, we restricted our laboratory analyses to two oxidative state markers: reactive oxygen metabolites (ROMs) concentration and plasma antioxidant capacity, following protocols adapted to small samples (Récapet et al. 2016c, 2017). Each sample was analysed on the same day in the laboratory for both ROM concentration and antioxidant capacity to avoid freeze-thaw cycles. The samples collected in different years were analysed in different years, so potential differences between years in ROM and OXY levels might partly reflect experimental variation and are thus not interpreted as possible biological effects later on. When studying correlations, these variables were however standardized within years, so that differences in their correlation coefficient could not stem from between-year differences.

Plasma concentration of ROMs was measured using the d-ROMs test (MC0001 kit, Diacron International, Grosseto, Italy). Haemolysed samples with a light orange or pink to bright red colouration were excluded visually, as well as opaque hyperlipemic samples with $OD_{800nm} > 0.100$ (N = 229, out of a total of 1116 samples). ROMs were measured on 5 to 7 different 96-well plates each year. The intra-plate repeatability [95% CI] was 0.797 [0.67; 0.897] on 24 duplicates and the inter-plate repeatability was 0.694 [0.579; 0.787] on 28 duplicates (calculated using the rptR package version 0.9.2; Nakagawa and Schielzeth 2010). Dietary hydroperoxides are degraded in the stomach (Kanazawa and Ashida 1998) and have a low uptake by intestinal cells (Maestre et al. 2013). Their contribution to plasma ROM concentration is most likely low. Plasmatic ROM concentration might however be influenced by the concentration of triglycerides in the plasma (Pérez-Rodríguez et al. 2015). Preliminary analyses on a subset of our samples provide no support for such association in our study species, and thus we did not control for the concentration of triglycerides in our statistical analyses (Appendix 1).

Plasma antioxidant capacity was measured by the capacity of plasma to oppose the oxidative action of the hypochlorous acid HClO (OXY adsorbent test, MC434 kit, Diacron International, Grosseto, Italy). This measure reflects the concentrations of ascorbate (vitamin C), flavonoids, carotenoids, glutathione and albumin, which are efficient scavengers of HClO, but not tocopherols (vitamin E) and ubiquinol, which are less reactive toward non-radical oxidants (Folkes et al. 1995; de Groot and Rauen 1998; Pattison et al. 2003; Pennathur et al. 2010). We chose this assay because it is less sensitive to variations in uric acid concentration in the plasma than other methods (Costantini 2011). ROMs were measured on 5 different 96-well plates each year. The intra-plate repeatability [bootstrap 95% CI] was 0.914 [0.876, 0.946] on 30 duplicates and the inter-plate repeatability was 0.858 [0.802, 0.901] on 24 duplicates.

Due to various technical problems during sample preparation, conservation and during laboratory assays, ROM concentration could not be measured for 11 out of 887 samples of adequate quality and plasma antioxidant capacity could not be measured for 27 out of 1116 samples. The total sample size was thus N = 876 for ROM concentration and N = 1089 for plasma antioxidant capacity. For correlation analyses, the sample size with data for both markers was N = 857. More detailed sample sizes by years, breeding stages, sexes and experimental treatments can be found in tables 1 to 4.

*Statistical analyses*

First, to assess the inter-annual repeatability of ROM and antioxidant capacity markers, we used samples collected on different years but at the same breeding stage (incubation or nestling feeding) for a given individual. We ran a linear mixed-effect model for each breeding stage because only females were sampled at the incubation stage whereas both males and females were sampled at the nestling feeding stage. We included individual identity (nested within sex) and plate (nested within year) as random factors, and year as a fixed factor. Individuals for which only one measure was available were included in the models to improve the estimates for residual variance (Martin et al. 2011). The effects of sex (two-level factor) and experimental manipulations (three-level factor: "control", "wing load manipulation", "food supplementation"), as well as their interaction, were included in the models describing the markers during nestling feeding. Second, to assess the repeatability of the oxidative state markers (ROM concentration and antioxidant capacity) measured at different breeding stages (incubation and nestling feeding) within the same year in a given female, we used linear mixed-effect models with individual identity within a year, individual identity within a breeding stage across years (to account for the potential correlation between measures of an individual at a given breeding stage between years), and plate (nested within year) as random factors, and year, breeding stage, manipulation and the interaction of breeding stage and manipulation as fixed factors, on females sampled during incubation and/or nestlings feeding. Repeatability was calculated as the ratio of individual (resp. individual within year) random variance on the sum of individual (resp. individual within year) and residual variances. Confidence interval for these estimates were calculated through 1000 parametric bootstraps and we tested whether the repeatability differed from zero through a likelihood-ratio test with a mixture distribution of Chi-square distributions with zero and one degree of freedom as reference, using the *rptR* package version 0.9.2 (Nakagawa and Schielzeth 2010).

Including body mass as a covariate in any of the models above did not change the variance estimates, and there was no significant effect of body mass, thus the models reported here do not include body mass. The parameters of the univariate models for the repeatability analyses were estimated by restricted maximum likelihood (REML) using the *lmer* function in R (Bates et al. 2014). The significance of the fixed effects was tested using F-tests with Satterthwaite estimation for the denominator degree of freedom, using the function *anova* from the *lmerTest* library (Kuznetsova et al. 2016).

To investigate the correlation between ROM concentration and antioxidant capacity, the two variables were modelled as response variables in bivariate mixed-effect models with plate as distinct random factors for antioxidant capacity and ROM concentration, as well individual identity as a common random factor when pooling multiple years. This allowed us to estimate the covariance, and thus correlation coefficients, at the between-individual and within-individual (residual) levels, while correcting for the random structure of both variables. The response variables were standardized to mean zero and variance one within each year, treatment and sex to account for potential differences in mean and variance between these groups. The covariance was estimated at the between- and within-individual levels when studying multiple years together. The total phenotypic variance-covariance matrix was computed as the sum of the between- and within-individual variance-covariance matrices. Only the phenotypic covariance is reported when there were less than 10 individuals with multiple measures.

The parameters of the bivariate models were estimated in a Bayesian framework that allowed us to fit different random effects for each response variable and to estimate their covariance at the between- and within-individual levels. The priors for the fixed effects estimates were set to a multinomial distribution with expected values of 0 and a diagonal variance-covariance matrix with a low strength of belief ($10^{10}$). The priors were set to inverse-Wishart distributions with the variances set to $1/n_i$ where $n_i$ was the number of variance components estimated for the parameter i, null covariances, and a degree of belief equal to the dimension of the variance-covariance matrix for the parameter. Preliminary analyses showed that the priors used for the covariances were quite informative on the posterior distribution of the between- and within-individual correlations, but not on the total phenotypic correlation. The analyses were performed with Markov chain Monte Carlo sampling using the *MCMCglmm* function in R (Hadfield 2010), with 1020000 iterations, a burn-in period of 20000, and a thinning interval of 500, to obtain autocorrelation values lower than 0.06 and an effective sample size higher than 2000 for all correlation estimates. We reported the mode of the posterior distribution as point estimate for the correlations and the Highest Posterior Density as 95% credibility interval. All estimates passed convergence diagnostic tests using the Cramer-von-Mises statistic with $P > 0.05$ (package CODA version 0.18-1; Plummer et al. 2006).

To test the effect of the wing load manipulation on the correlation between the two markers, we compared two bivariate models describing the standardized markers in 2012 and 2013, one with homogeneous within-individual (residual) covariances according to the experimental treatment and a second allowing for heterogeneous within-individual covariances according to the experimental treatment. Similarly, we tested the effect of the food supplementation on the correlation by comparing bivariate models for the standardized markers in 2014 with or without heterogeneous covariances according to the experimental treatment. Finally, the effect of temporal variation in the environment was tested by comparing two models for the control groups in all years, with or without heterogeneous covariances according to year. The Deviance Information Criterion was used as an indication to compare models with different random variance structures (DIC; Spiegelhalter et al. 2002), with a DIC difference larger than five interpreted as a significantly better model.

# Results

*Inter-annual repeatability of OXY and ROM*

In females sampled during incubation in different years, the repeatability of antioxidant capacity was significant (*r* [95% confidence interval] = 0.581 [0.327; 0.825]; Table 1). In contrast, the repeatability of ROM concentration was very low and did not significantly differ from zero (0.032 [0; 0.660]; Table 1).

In males and females sampled during feeding in different years, the repeatability of antioxidant capacity was low but significant (0.124 [0.018; 0.254]; Table 1), whereas the repeatability of ROM concentration did not significantly differ from zero (0.061 [0; 0.257]; Table 1). There was no effect of the manipulations, even in interaction with sex, on antioxidant capacity (manipulations x sex = $F_{2,796}$ = 0.08, P = 0.92; manipulations: $F_{2,812}$ = 1.55, P = 0.21) or ROM concentration (manipulations x sex = $F_{2,640}$ = 0.17, P = 0.84; manipulations: $F_{2,650}$ = 0.41, P = 0.67). During feeding, antioxidant capacity was independent of sex ($F_{1,646}$ = 0.39, P = 0.54) but ROM concentration was lower in males than females (-0.095 ± 0.045, $F_{1,512}$ = 12.34, P = 0.0005).

*Intra-annual repeatability of OXY and ROM*

In females sampled at different breeding stages within the same year, the repeatability of antioxidant capacity was low but significant (0.131 [0.030; 0.253]; Table 1), whereas the repeatability of ROM concentration between stages was null and non-significant (<0.001 [0; 0.174]; Table 1). The food supplementation had a positive effect on the increase in antioxidant capacity between the incubation and breeding stage (interaction breeding stage x manipulations: $F_{2,293}$ = 3.63, P = 0.028, Figure 1a). There was no effect on ROM concentration of the manipulations, either alone ($F_{2,519}$ = 1.08, P = 0.34) or in interaction with the breeding stage ($F_{2,429}$ = 0.14, P = 0.87). ROM concentration was higher during the nestling feeding stage compared to the incubation stage (+0.165 ± 0.058, $F_{1,100}$ = 13.14, P = 0.0005, Figure 1b).

*Correlations between physiological markers*

Overall, there was a positive correlation between ROM concentration and antioxidant capacity during nestling feeding at the phenotypic level (N = 527 individuals; n = 646 observations; posterior mode [95% credibility interval]: $V_{between-individual}$ = 0.242 [-0.271; 0.618], $V_{within-individual}$ = 0.099 [-0.043; 0.193], $V_{phenotypic}$ = 0.102 [0.028; 0.181]). The strength of this phenotypic correlation however differed according to the experimental group: the correlation was stronger in wing load manipulated (handicapped) females compared to control ones (model with heterogeneous covariances according to manipulation compared to homogeneous covariances: ΔDIC = -12.6; Table 2), whereas there was no difference in correlation between males whose females were handicapped and controls (ΔDIC = -4.0; Table 2). Conversely, there was no effect of the food supplementation in 2014 on the correlation between antioxidant capacity and ROM concentration, since the positive correlation found in supplemented pairs was similar to that in control ones (model with heterogeneous covariances according to food supplementation compared to homogeneous covariances: ΔDIC = +4.6; Table 3). Finally, in control pairs, the strength of the correlation was also higher in 2014 compared to 2012 and 2013 (model with heterogeneous covariances according to year compared to homogeneous covariances: ΔDIC = -26.3; Table 4). Evidence for a positive correlation between markers was weaker in incubating females compared to nestling-feeding adults (Table 4).

## Discussion

In this study, we aimed at describing the correlation structure between two plasmatic markers of the oxidative balance and its variation in different natural and experimental conditions.

Individual identity and food availability were important determinants of antioxidant capacity but did not influence ROM concentration. These two physiological markers covaried positively. Importantly, the correlation between markers was stronger in experimentally handicapped (wing-load manipulated) females and varied between years, while relieving energetic constraints through food supplementation did not change the correlation. Overall, our findings refute the existence of a stable correlation structure between these two widely used markers, and call for caution when interpreting these, or similar, markers of oxidative stress.

*Stability and variation of individual markers*

Antioxidant capacity was repeatable between years at a given breeding stage, as well as within year between incubation and nestling feeding stages (for females). The repeatability was higher during incubation compared to nestling feeding (between years) or between stages (within the same year), suggesting that uncontrolled environmental factors had a stronger effect during nestling feeding. It thus seems that the between-year variation in antioxidant capacity is partly determined by individual characteristics, perhaps including the ability to find higher quality food rich in antioxidants, to acquire a better territory, or to produce more enzymatic antioxidants and spare the dietary antioxidant pools. This could be due to genetic differences or to permanent individual differences due to early-life effects. Similarly, significant within-individual repeatability in antioxidant capacity among breeding seasons was found in barn swallows *Hirundo rustica* (r = 0.487, P < 0.001, Saino *et al.* 2011), Seychelles warblers *Acrocephalus sechellensis* (r = 0.122, P = 0.043, van de Crommenacker et al. 2011) and European shags *Phalacrocorax aristotelis*, although repeatability varied with age (2-9 years old: *r* = 0.20, P = 0.36; 10 to 22 years old: *r* = 0.33, P = 0.020; Herborn *et al.* 2015). The lower repeatability at the nestling feeding stage, or between breeding stages within the same year, might be partly explained by individual variation in the response to environmental conditions, such as the increase in antioxidant capacity in food-supplemented females between incubation and nestling feeding. Previous experimental studies that manipulated energy expenditure or oxidative balance in birds showed that measures of antioxidant capacity were only repeatable in control birds (Meitern et al. 2013) or were not correlated across different treatments (Beamonte-Barrientos and Verhulst 2013).

ROM concentration was not repeatable either between- or within-year; variations in ROM concentration were thus mainly determined by external factors or physiological changes, apart from a difference between males and females. The lower methodological reliability of ROM measurements compared to antioxidant capacity, mainly due to low values of ROMs levels, could also have reduced our ability to detect weak but biologically significant repeatability. This result here contrasts with studies of ROMs that found individual consistency even under different experimental treatments (Stier *et al*. 2012; Beamonte-Barrientos & Verhulst 2013; Herborn *et al*. 2015; but see van de Crommenacker *et al*. 2011). Individual consistency in ROM measurements can however vary with age (Herborn et al. 2016). Contrary to the positive effect of food supplementation on antioxidant capacity, ROM concentration was not influenced by the experimental manipulations of energetic constraints (wing load manipulation or food supplementation). These results contrast with the finding that food supplementation decreased ROM concentration but had no effect on antioxidant capacity in breeding great tit females (Giordano et al. 2015). In our study, ROM concentration strongly increased between incubation and the nestling feeding period in females. The increase between breeding stages in females might reflect the costs of reproduction: ROM concentration increased in breeding females, but not in non-breeding ones, in mice (Stier et al. 2012) and in breeding female Seychelles warblers naturally infected with malaria (van de Crommenacker et al. 2012). Alternatively oxidative damage might be adaptively reduced in females during egg-laying to avoid negative effects on offspring, a process called oxidative shielding (Giordano et al. 2015; Blount et al. 2016).

*A highly variable correlation structure*

Antioxidant capacity was positively correlated with ROM concentration when considering all individuals irrespective of their experimental treatment. This positive correlation could not be fully explained by individual differences in quality or resource acquisition (van Noordwijk and de Jong 1986; Wilson and Nussey 2009) as it was also found at the within-individual level. It is thus inconsistent with the hypothesis that circulating antioxidants are limited by external factors and can be depleted when protecting individuals against increased reactive oxygen species. Indeed, in such a case, antioxidant defences and ROMs should correlate negatively or be modified in opposite ways depending on individual and environmental factors (Fletcher et al. 2013; Yang et al. 2013; Hanssen et al. 2013; López-Arrabé et al. 2014, 2015). Positive correlations or variation in the same direction between oxidative damage and antioxidant capacity were also found in previous studies on birds (van de Crommenacker et al. 2011; Stier et al. 2012; Isaksson 2013; Xu et al. 2014; Beaulieu et al. 2015; Vaugoyeau et al. 2015; Marasco et al. 2017). Antioxidant protection could thus adaptively build up to face increased exposure to ROS in periods of higher energy demands, especially in females during the early breeding stages, when oxidative stress can be particularly harmful for developing offspring (Blount et al. 2016).

The correlation between ROM concentration and antioxidant capacity was sensitive to variation in the conditions experienced by individuals. It was stronger in handicapped females, which experienced stronger energetic constraints compared to control ones. Conversely, experimentally increased food availability did not appear to influence the correlation between the markers, as there was no noticeable difference in this correlation between control and food-supplemented pairs. The strength of the correlation also increased in control birds from 2012 to 2014. This hints at a role of environmental conditions in modulating the correlation between antioxidant defences and oxidative damages. Indeed, particularly sunny and dry meteorological conditions in 2012 and particularly rainy and cold conditions in 2014 resulted in strong differences in mortality rate among nestlings and thus in reproductive performances, with 2013 in between (mean fledging success in control nests ± S.D. (N) = 4.7 ± 2.6 (89) in 2012, 3.0 ± 2.4 (99) in 2013, and 1.5 ± 2.0 (82) in 2014). The correlation between oxidative damage and antioxidant capacity might thus be stronger when environmental conditions are harsher, although this cannot be formally tested here with only three study years. Overall, the correlation appeared more salient when individuals were energetically constrained either experimentally or naturally. This was not an artefact of a larger range of values for the markers (heterogeneity in individual responses) in more constrained habitats, as variances for ROM concentration and plasma antioxidant capacity actually tended to be larger in control females than in handicapped females, and in 2012 and 2013 than in 2014 (data not shown). Although the exact sources of the observed differences in the correlation remain speculative, our results clearly show that the relationships between different components of the oxidative balance are not fixed but may be modulated by individual or environmental factors.

Despite our limited understanding of oxidative stress and redox signalling across animal species (Halliwell and Gutteridge 2015; Jones and Sies 2015), a stable relationship between oxidative damage and antioxidant protection is often assumed: it underpins some proposed integrative measures of oxidative stress, such as the ratio between a marker of ROS production/oxidative damages and a marker of antioxidant protection (e.g. Costantini et al. 2006), or the extraction of principal components from PCAs on a set of these markers (Hõrak and Cohen 2010). Our results however call for more caution when modelling the relationship between markers of oxidative damages and antioxidants. Consistently, previous studies failed to find consistent correlations between multiple markers of oxidative balance, despite large longitudinal samples (Romero-Haro and Alonso-Alvarez 2014; Christensen et al. 2015). Correlations between antioxidants alone were also found to vary strikingly among bird species (Cohen and McGraw 2009). Beyond defining reference values for different environmental conditions and life-stages (Beaulieu and Costantini 2014), the stability of the correlation structure under these varying conditions should thus be tested in any study species before making inferences based on ratios or principal components analyses. Such stability may however not exist in nature and more mechanistic models of oxidative homeostasis may thus be required to properly measure oxidative stress. This type of mechanistic models is now relatively well implemented and widely used for studying energetic trade-offs, building upon thermodynamics and chemistry principles (Jusup et al. 2017). These models already include nutritional trade-offs due to dietary restrictions in some nutrients, especially in plants. Other potential sources of trade-offs, such as oxidative stress, but also immune function or glucose homeostasis (Récapet et al. 2016b; Montoya et al. 2018), are however not modelled explicitly, probably due to the complexity of their interactions (Cohen et al. 2012). Theoretical models have however shown that multiple, partially independent physiological mechanisms underlying life-history trade-offs could relax the overall trade-off observed at the individual level and thus change the evolutionary outcome (Cohen et al. 2017). The physiological "machinery" evolved to respond to environmental challenges can be seen as an evolutionary constraint that does matter and should thus be taken into account when projecting the future evolution of organisms.

## Conclusions

In our study, individual antioxidant capacity was repeatable within and between years, but ROM concentration was not, which suggest that an individual's ability to acquire antioxidant-rich food plays an important role in its ability to respond to oxidative stress. The two markers were positively correlated and this relationship was conditional on the energetic constraints experienced by each individual: correlations between markers were stronger when wing load was increased experimentally or when environmental conditions were naturally poorer. This probably comes from a tighter adjustment of antioxidant defences to ROS production when conditions were constrained, which could be particularly important to maintain cell redox homeostasis. Given the diversity of physiological and ecological factors that seem to affect the correlation structure between markers of oxidative state in a single population, the discrepancies in these correlations among studies, species and populations come as no surprise. Our results thus question our ability to interpret and make ecological inferences from markers of oxidative state without a more flexible, mechanistic understanding of their interactions.

**Compliance with Ethical Standards**

The authors declare that they have no conflict of interest.

All applicable international, national, and/or institutional guidelines for the care and use of animals were followed. Permission for catching and ringing adult and young birds was granted by the Ringing Centre from the Museum of Natural History in Stockholm (license number 471:M009 to C.R.). Permission for blood taking and experimental procedures was granted by the Ethical Committee for Experiments on Animals in Sweden (license number C 108/7).

**Data availability**

The datasets and as well as the results of the MCMC sampling generated during the current study are available in the figshare repository, [PERSISTENT WEB LINK TO DATASETS].

# References

Ardia DR, Clotfelter ED (2007) Individual quality and age affect responses to an energetic constraint in a cavity-nesting bird. Behav Ecol 18:259–266. doi: 10.1093/beheco/arl078

Avery S V (2014) Oxidative stress and cell function. In: Laher I (ed) Systems Biology of Free Radicals and Antioxidants. Springer-Verlag, Berlin Heidelberg, pp 89–112

Barja G (2007) Mitochondrial oxygen consumption and reactive oxygen species production are independently modulated: Implications for aging studies. Rejuvenation Res 10:215–224. doi: 10.1089/rej.2006.0516

Bates D, Maechler M, Bolker B, Walker S (2014) lme4: Linear mixed-effects models using Eigen and S4. R package version 1.1-7

Beamonte-Barrientos R, Verhulst S (2013) Plasma reactive oxygen metabolites and non-enzymatic antioxidant capacity are not affected by an acute increase of metabolic rate in zebra finches. J Comp Physiol B 183:675–683. doi: 10.1007/s00360-013-0745-4

Beaulieu M, Costantini D (2014) Biomarkers of oxidative status: missing tools in conservation physiology. Conserv Physiol 2:cou014. doi: 10.1093/conphys/cou014

Beaulieu M, Geiger RE, Reim E, et al (2015) Reproduction alters oxidative status when it is traded-off against longevity. Evolution (N Y) 69:1786–1796. doi: 10.1111/evo.12697

Beaulieu M, Reichert S, Le Maho Y, et al (2011) Oxidative status and telomere length in a long-lived bird facing a costly reproductive event. Funct Ecol 25:577–585. doi: 10.1111/j.1365-2435.2010.01825.x

Beaulieu M, Schaefer HM (2014) The proper time for antioxidant consumption. Physiol Behav 128C:54–59. doi: 10.1016/j.physbeh.2014.01.035

Blount JD, Vitikainen EIK, Stott I, Cant MA (2016) Oxidative shielding and the cost of reproduction. Biol Rev 91:483–497. doi: 10.1111/brv.12179

Bonawitz ND, Chatenay-Lapointe M, Pan Y, Shadel GS (2007) Reduced TOR signaling extends chronological life span via increased respiration and upregulation of mitochondrial gene expression. Cell Metab 5:265–277. doi: 10.1016/j.cmet.2007.02.009

Brand MD (2000) Uncoupling to survive? The role of mitochondrial inefficiency in ageing. Exp Gerontol 35:811–820. doi: 10.1016/S0531-5565(00)00135-2

Christensen LL, Selman C, Blount JD, et al (2015) Plasma markers of oxidative stress are uncorrelated in a wild mammal. Ecol Evol 5:5096–5108. doi: 10.1002/ece3.1771

Cohen AA, Isaksson C, Salguero-Gómez R (2017) Co-existence of multiple trade-off currencies shapes evolutionary outcomes. PLoS One 12:e0189124. doi: 10.1371/journal.pone.0189124

Cohen AA, Martin LB, Wingfield JC, et al (2012) Physiological regulatory networks: ecological roles and evolutionary constraints. Trends Ecol Evol 27:428–435. doi: 10.1016/j.tree.2012.04.008

Cohen AA, McGraw KJ (2009) No simple measures for antioxidant status in birds: Complexity in inter- and intraspecific correlations among circulating antioxidant types. Funct Ecol 23:310–320. doi: 10.1111/j.1365-2435.2009.01540.x

Costantini D (2011) On the measurement of circulating antioxidant capacity and the nightmare of uric acid. Methods Ecol Evol 2:321–325. doi: 10.1111/j.2041-210X.2010.00080.x

Costantini D, Casagrande S, De Filippis S, et al (2006) Correlates of oxidative stress in wild kestrel nestlings (Falco tinnunculus). J Comp Physiol B 176:329–337. doi: 10.1007/s00360-005-0055-6

Costantini D, Verhulst S (2009) Does high antioxidant capacity indicate low oxidative stress? Funct Ecol 23:506–509. doi: 10.1111/j.1365-2435.2009.01546.x

Criscuolo F, Gonzalez-Barroso M del M, Bouillaud F, et al (2005) Mitochondrial uncoupling

proteins: New perspectives for evolutionary ecologists. Am Nat 166:686–699. doi: 10.1086/497439

de Groot H, Rauen U (1998) Tissue injury by reactive oxygen species and the protective effects of flavonoids. Fundam Clin Pharmacol 12:249–255. doi: 10.1111/j.1472-8206.1998.tb00951.x

Fletcher QE, Selman C, Boutin S, et al (2013) Oxidative damage increases with reproductive energy expenditure and is reduced by food-supplementation. Evolution 67:1527–1536. doi: 10.1111/evo.12014

Folkes LK, Candeias LP, Wardman P (1995) Kinetics and mechanisms of hypochlorous acid reactions. Arch Biochem Biophys 323:120–126. doi: 10.1006/abbi.1995.0017

Geiger S, Le Vaillant M, Lebard T, et al (2012) Catching-up but telomere loss: half-opening the black box of growth and ageing trade-off in wild king penguin chicks. Mol Ecol 21:1500–1510. doi: 10.1111/j.1365-294X.2011.05331.x

Gems D, Partridge L (2013) Genetics of longevity in model organisms: Debates and paradigm shifts. Annu Rev Physiol 75:621–644. doi: 10.1146/annurev-physiol-030212-183712

Giordano M, Costantini D, Pick JL, Tschirren B (2015) Female oxidative status, egg antioxidant protection and eggshell pigmentation: a supplemental feeding experiment in great tits. Behav Ecol Sociobiol 69:777–785. doi: 10.1007/s00265-015-1893-1

Glazier DS (2015) Is metabolic rate a universal 'pacemaker' for biological processes? Biol Rev 90:377–407. doi: 10.1111/brv.12115

Hadfield JD (2010) MCMC methods for multi-response generalized linear mixed models: The MCMCglmm R package. J Stat Softw 33:1–22. doi: 10.18637/jss.v033.i02

Halliwell B, Gutteridge JMC (2015) Free Radicals in Biology and Medicine, 5th editio. Oxford University Press

Hanssen SA, Bustnes JO, Schnug L, et al (2013) Antiparasite treatments reduce humoral immunity and impact oxidative status in raptor nestlings. Ecol Evol 3:5157–5166. doi: 10.1002/ece3.891

Herborn KA, Daunt F, Heidinger BJ, et al (2016) Age, oxidative stress exposure and fitness in a long-lived seabird. Funct Ecol 30:913–921. doi: 10.1111/1365-2435.12578

Hõrak P, Cohen AA (2010) How to measure oxidative stress in an ecological context: methodological and statistical issues. Funct Ecol 24:960–970. doi: 10.1111/j.1365-2435.2010.01755.x

Isaksson C (2013) Opposing effects on glutathione and reactive oxygen metabolites of sex, habitat, and spring date, but no effect of increased breeding density in great tits (Parus major). Ecol Evol 3:2730–2738. doi: 10.1002/ece3.663

Isaksson C, Sheldon BC, Uller T (2011) The challenges of integrating oxidative stress into life-history biology. Bioscience 61:194–202. doi: 10.1525/bio.2011.61.3.5

Jones DP, Sies H (2015) The redox code. Antioxid Redox Signal 23:734–746. doi: 10.1089/ars.2015.6247

Jusup M, Sousa T, Domingos T, et al (2017) Physics of metabolic organization. Phys Life Rev 20:1–39. doi: 10.1016/j.plrev.2016.09.001

Kanazawa K, Ashida H (1998) Dietary hydroperoxides of linoleic acid decompose to aldehydes in stomach before being absorbed into the body. Biochim Biophys Acta - Lipids Lipid Metab 1393:349–361. doi: 10.1016/S0005-2760(98)00089-7

Kuznetsova A, Brockhoff PB, Christensen RHB (2016) lmerTest: Tests in Linear Mixed Effects Models. R package version 2.0-32

López-Arrabé J, Cantarero A, Pérez-Rodríguez L, et al (2015) Nest-dwelling ectoparasites reduce antioxidant defences in females and nestlings of a passerine: a field experiment.

Oecologia 179:29–41. doi: 10.1007/s00442-015-3321-7

López-Arrabé J, Cantarero A, Pérez-Rodríguez L, et al (2014) Plumage ornaments and reproductive investment in relation to oxidative status in the Iberian Pied Flycatcher (Ficedula hypoleuca iberiae). Can J Zool 92:1019–1027. doi: 10.1139/cjz-2014-0199

Maestre R, Douglass JD, Kodukula S, et al (2013) Alterations in the intestinal assimilation of oxidized PUFAs are ameliorated by a polyphenol-rich grape seed extract in an in vitro model and Caco-2 cells. J Nutr 143:295–301. doi: 10.3945/jn.112.160101

Marasco V, Stier A, Boner W, et al (2017) Environmental conditions can modulate the links among oxidative stress, age, and longevity. Mech Ageing Dev 164:100–107. doi: 10.1016/j.mad.2017.04.012

Markó G, Costantini D, Michl G, Török J (2011) Oxidative damage and plasma antioxidant capacity in relation to body size, age, male sexual traits and female reproductive performance in the collared flycatcher (Ficedula albicollis). J Comp Physiol B 181:73–81. doi: 10.1007/s00360-010-0502-x

Martin JGA, Nussey DH, Wilson AJ, Réale D (2011) Measuring individual differences in reaction norms in field and experimental studies: A power analysis of random regression models. Methods Ecol Evol 2:362–374. doi: 10.1111/j.2041-210X.2010.00084.x

Meitern R, Sild E, Kilk K, et al (2013) On the methodological limitations of detecting oxidative stress: effects of paraquat on measures of oxidative status in greenfinches. J Exp Biol 216:2713–2721. doi: 10.1242/jeb.087528

Metcalfe NB, Alonso-Alvarez C (2010) Oxidative stress as a life-history constraint: the role of reactive oxygen species in shaping phenotypes from conception to death. Funct Ecol 24:984–996. doi: 10.1111/j.1365-2435.2010.01750.x

Monaghan P, Metcalfe NB, Torres R (2009) Oxidative stress as a mediator of life history trade-

offs: mechanisms, measurements and interpretation. Ecol Lett 12:75–92. doi: 10.1111/j.1461-0248.2008.01258.x

Montoya B, Briga M, Jimeno B, et al (2018) Baseline glucose level is an individual trait that is negatively associated with lifespan and increases due to adverse environmental conditions during development and adulthood. J Comp Physiol B 188:517–526. doi: 10.1007/s00360-017-1143-0

Moreno J, Merino S, Potti J, et al (1999) Maternal energy expenditure does not change with flight costs or food availability in the pied flycatcher ( Ficedula hypoleuca ): costs and benefits for nestlings. Behav Ecol Sociobiol 46:244–251. doi: 10.1007/s002650050616

Nakagawa S, Schielzeth H (2010) Repeatability for Gaussian and non-Gaussian data: a practical guide for biologists. Biol Rev Camb Philos Soc 85:935–956. doi: 10.1111/j.1469-185X.2010.00141.x

Nikolaidis MG, Jamurtas AZ, Paschalis V, et al (2008) The effect of muscle-damaging exercise on blood and skeletal muscle oxidative stress: Magnitude and time-course considerations. Sport Med 38:579–606. doi: 10.2165/00007256-200838070-00005

Pattison DI, Hawkins CL, Davies MJ (2003) Hypochlorous acid-mediated oxidation of lipid components and antioxidants present in low-density lipoproteins: absolute rate constants for the reaction of hypochlorous acid with protein side chains and peptide bonds. Chem Res Toxicol 16:439–449. doi: 10.1021/tx025670s

Pennathur S, Maitra D, Byun J, et al (2010) Potent antioxidative activity of lycopene: A potential role in scavenging hypochlorous acid. Free Radic Biol Med 49:205–213. doi: 10.1016/j.freeradbiomed.2010.04.003

Pérez-Rodríguez L, Romero-Haro AA, Sternalski A, et al (2015) Measuring oxidative stress: The confounding effect of lipid concentration in measures of lipid peroxidation. Physiol

Biochem Zool 88:345–351. doi: 10.1086/680688

Plummer M, Best N, Cowles K, Vines K (2006) CODA: Convergence Diagnosis and Output Analysis for MCMC. R News 6:7–11

Récapet C, Bize P, Doligez B (2017) Food availability modulates differences in parental effort between dispersing and philopatric birds. Behav Ecol 28:688–697. doi: 10.1093/beheco/arx017

Récapet C, Daniel G, Taroni J, et al (2016a) Food supplementation mitigates dispersal-dependent differences in nest defence in a passerine bird. Biol Lett 12:20160097. doi: 10.1098/rsbl.2016.0097

Récapet C, Sibeaux A, Cauchard L, et al (2016b) Selective disappearance of individuals with high levels of glycated haemoglobin in a free-living bird. Biol Lett 12:20160243. doi: 10.1098/rsbl.2016.0243

Récapet C, Zahariev A, Blanc S, et al (2016c) Differences in the oxidative balance of dispersing and non-dispersing individuals: an experimental approach in a passerine bird. BMC Evol Biol 16:125. doi: 10.1186/s12862-016-0697-x

Reichert S, Stier A, Zahn S, et al (2014) Increased brood size leads to persistent eroded telomeres. Front Ecol Evol 2:1–11. doi: 10.3389/fevo.2014.00009

Romero-Haro AA, Alonso-Alvarez C (2014) Covariation in oxidative stress markers in the blood of nestling and adult birds. Physiol Biochem Zool 87:353–362. doi: 10.1086/674432

Saino N, Caprioli M, Romano M, et al (2011) Antioxidant defenses predict long-term survival in a passerine bird. PLoS One 6:e19593. doi: 10.1371/journal.pone.0019593

Salin K, Auer SK, Rey B, et al (2015a) Variation in the link between oxygen consumption and ATP production, and its relevance for animal performance. Proc R Soc B Biol Sci 282:20151028. doi: 10.1098/rspb.2015.1028

Salin K, Auer SK, Rudolf AM, et al (2015b) Individuals with higher metabolic rates have lower levels of reactive oxygen species in vivo. Biol Lett 11:20150538. doi: 10.1098/rsbl.2015.0538

Sanz JJ, Kranenbarg S, Tinbergen JM (2000) Differential response by males and females to manipulation of partner contribution in the great tit (Parus major). J Anim Ecol 69:74–84. doi: 10.1046/j.1365-2656.2000.00373.x

Sena LA, Chandel NS (2012) Physiological roles of mitochondrial reactive oxygen species. Mol Cell 48:158–166. doi: 10.1016/j.molcel.2012.09.025

Speakman JR, Krol E (2010) The heat dissipation limit theory and evolution of life histories in endotherms-Time to dispose of the disposable soma theory? Integr Comp Biol 50:793–807. doi: 10.1093/icb/icq049

Speakman JR, Selman C (2011) The free-radical damage theory: Accumulating evidence against a simple link of oxidative stress to ageing and lifespan. Bioessays 33:255–259. doi: 10.1002/bies.201000132

Spiegelhalter DJ, Best NG, Carlin BP, van der Linde A (2002) Bayesian measures of model complexity and fit. J R Stat Soc Ser B 64:583–639. doi: 10.1111/1467-9868.00353

Stearns SC (1989) Trade-offs in life-history evolution. Funct Ecol 3:259–268. doi: 10.2307/2389364

Stier A, Bize P, Habold C, et al (2014a) Mitochondrial uncoupling prevents cold-induced oxidative stress: a case study using UCP1 knockout mice. J Exp Biol 217:624–630. doi: 10.1242/jeb.092700

Stier A, Massemin S, Criscuolo F (2014b) Chronic mitochondrial uncoupling treatment prevents acute cold-induced oxidative stress in birds. J Comp Physiol B 184:1021–1029. doi: 10.1007/s00360-014-0856-6

Stier A, Reichert S, Massemin S, et al (2012) Constraint and cost of oxidative stress on reproduction: correlative evidence in laboratory mice and review of the literature. Front Zool 9:37. doi: 10.1186/1742-9994-9-37

van de Crommenacker J, Komdeur J, Burke T, Richardson DS (2011) Spatio-temporal variation in territory quality and oxidative status: a natural experiment in the Seychelles warbler (Acrocephalus sechellensis). J Anim Ecol 80:668–680. doi: 10.1111/j.1365-2656.2010.01792.x

van de Crommenacker J, Richardson DS, Koltz a. M, et al (2012) Parasitic infection and oxidative status are associated and vary with breeding activity in the Seychelles warbler. Proc R Soc B Biol Sci 279:1466–1476. doi: 10.1098/rspb.2011.1865

van Noordwijk AJ, de Jong G (1986) Acquisition and allocation of resources: Their influence on variation in life history tactics. Am Nat 128:137–142. doi: 10.1086/284547

Vaugoyeau M, Decenciere B, Perret S, et al (2015) Is oxidative status influenced by dietary carotenoid and physical activity after moult in the great tit (Parus major)? J Exp Biol 218:2106–2115. doi: 10.1242/jeb.111039

Veskoukis AS, Nikolaidis MG, Kyparos A, Kouretas D (2009) Blood reflects tissue oxidative stress depending on biomarker and tissue studied. Free Radic Biol Med 47:1371–1374. doi: 10.1016/j.freeradbiomed.2009.07.014

Wegmann M, Voegeli B, Richner H (2015) Physiological responses to increased brood size and ectoparasite infestation: Adult great tits favour self-maintenance. Physiol Behav 141:127–134. doi: 10.1016/j.physbeh.2015.01.017

Wilson AJ, Nussey DH (2009) What is individual quality? An evolutionary perspective. Trends Ecol Evol 25:207–214. doi: 10.1016/j.tree.2009.10.002

Xu Y-C, Yang D-B, Speakman JR, Wang D-H (2014) Oxidative stress in response to natural and experimentally elevated reproductive effort is tissue dependent. Funct Ecol 28:402–410. doi: 10.1111/1365-2435.12168

Yang D-B, Xu Y-C, Wang D-H, Speakman JR (2013) Effects of reproduction on immuno-suppression and oxidative damage, and hence support or otherwise for their roles as mechanisms underpinning life history trade-offs, are tissue and assay dependent. J Exp Biol 216:4242–4250. doi: 10.1242/jeb.092049

**Fig. 1 Antioxidant capacity (a) and log-transformed ROM concentration (b) in female collared flycatchers according to experimental treatment and breeding stage (incubation = pre-treatment, feeding = post-treatment).** Values were corrected for the year effect.

**Table 1: Repeatability of the measures of reactive oxygen metabolites (ROMs) and antioxidant capacity (OXY test) at different time scales and different breeding stages.** Estimates are obtained from linear-mixed models, 95% confidence interval from parametric bootstraps and P values from likelihood-ratio tests. The number of individuals is given in parentheses below the number of observations.

| Repeatability | Sex | Variable | N | r | CI$_{95\%}$ | P |
|---|---|---|---|---|---|---|
| Between years during incubation | F | OXY | 251 (234) | **0.581** | **[0.327; 0.825]** | **0.02** |
| Between years during incubation | F | ROMs | 214 (202) | 0.032 | [0; 0.660] | 0.44 |
| Between years during nestling feeding | F & M | OXY | 838 (652) | **0.124** | **[0.018; 0.254]** | **0.02** |
| Between years during nestling feeding | F & M | ROMs | 663 (535) | 0.061 | [0; 0.257] | 0.24 |
| Between breeding stages within a year | F | OXY | 681 (359) | **0.131** | **[0.030; 0.253]** | **0.02** |
| Between breeding stages within a year | F | ROMs | 535 (310) | 0.000 | [0; 0.174] | 0.50 |

**Table 2: Effect of the wing load manipulation in 2012 and 2013 on the correlation between reactive oxygen metabolites (ROMs) and antioxidant capacity (OXY test).** Mode and 95% credibility interval of the posterior distribution of the coefficients of correlation between ROM concentration and antioxidant capacity, for different experimental groups, sexes and years. The number of individuals is given in parentheses below the sample sizes when the samples covered multiple years, and thus some individuals were sampled more than once. When 10 or more individual birds were measured repeatedly, the within-individual correlation (residual correlation) is reported separately from the between-individual correlation.

| Treatment | Sex | N | DIC | Between-individual correlation | Within-individual correlation | Total phenotypic correlation |
|---|---|---|---|---|---|---|
| Both groups, with the same correlation coefficients | F | 187 (176) | **982.4** | 0.143 [-0.477; 0.729] | 0.088 [-0.177; 0.376] | 0.099 [-0.045; 0.245] |
| Both groups, with different correlation coefficients | F | 187 (176) | **969.9** | 0.279 [-0.364; 0.731] | 0.151 [-0.273; 0.475] in manipulated females | |
| | | | | | -0.083 [-0.368; 0.275] in control females | |
| Wing load manipulation | F | 91 (87) | | - | - | 0.168 [-0.038; 0.374] |
| Control | F | 96 (93) | | - | - | -0.004 [-0.203; 0.186] |
| Both groups, with the same correlation coefficients | M | 220 (191) | 1195.6 | 0.337 [-0.343; 0.714] | 0.028 [-0.172; 0.217] | 0.046 [-0.075; 0.189] |
| Both groups, with different correlation coefficients | M | 220 (191) | 1191.6 | 0.320 [-0.342; 0.726] | -0.065 [-0.347; 0.171] | in males paired with manipulated females |
| | | | | | 0.104 [-0.163; 0.360] | in control males |
| Wing load manipulation | M | 120 (112) | | - | - | 0.028 [-0.165; 0.202] |
| Control | M | 100 (94) | | - | - | 0.138 [-0.062; 0.331] |

**Table 3: Effect of the food supplementation in 2014 on the correlation between reactive oxygen metabolites (ROMs) and antioxidant capacity (OXY test) with both sexes pooled.** Mode and 95% credibility interval of the posterior distribution of the coefficients of correlation between ROM concentration and antioxidant capacity for males and females in 2014.

| Treatment | N | DIC | Phenotypic correlation | |
|---|---|---|---|---|
| Both groups, with the same correlation coefficients | 239 | 1283.5 | **0.169 [0.050; 0.290]** | |
| Both groups, with different correlation coefficients | 239 | 1288.0 | **0.181 [0.006; 0.339]** | in the food supplemented group |
| | | | 0.148 [-0.055; 0.313] | in the control group |
| Food supplementation | 128 | | **0.186 [0.010; 0.342]** | |
| Control | 111 | | 0.133 [-0.065; 0.307] | |

**Table 4: Temporal and seasonal variation in the correlation between reactive oxygen metabolites (ROMs) and antioxidant capacity (OXY test).** Mode and 95% credibility interval of the posterior distribution of the coefficients of correlation between ROM concentration and antioxidant capacity, in nestling-feeding males and females from control groups and in incubating females before the start of the treatments. The number of individuals is given in parentheses below the sample sizes when the samples covered multiple years, and thus some individuals were sampled more than once. When 10 or more individual birds were measured repeatedly, the within-individual correlation (residual correlation) is reported separately from the between-individual correlation.

| Sex and stage | Years | N | DIC | Between-individual | Within-individual | | Total phenotypic |
|---|---|---|---|---|---|---|---|
| F (incubating) | 2012-2014 | 211 (199) | | -0.101 [-0.822; 0.691] | 0.050 [-0.175; 0.298] | | 0.049 [-0.090; 0.190] |
| F + M (feeding) | 2012-2014 | 307 (274) | **1646.3** | 0.042 [-0.637; 0.482] | 0.137 [-0.066; 0.320] | | 0.104 [-0.019; 0.200] |
| F + M (feeding) | 2012-2014 | 307 (274) | **1620.1** | -0.114 [-0.585; 0.549] | 0.087 [-0.277; 0.445] | in 2012 | |
| | | | | | 0.114 [-0.171; 0.373] | in 2013 | |
| | | | | | 0.262 [-0.083; 0.545] | in 2014 | |
| F + M (feeding) | 2012 | 72 | | - | - | | 0.064 [-0.180; 0.282] |
| F + M (feeding) | 2013 | 124 | | - | - | | 0.063 [-0.096; 0.251] |
| F + M (feeding) | 2014 | 111 | | - | - | | 0.132 [-0.028; 0.325] |

**Appendix 1: Correlation between markers of oxidative balance and nutritional state in the plasma.**

In 2014, the concentrations of triglycerides, glucose and lactate were measured in whole blood immediately after blood taking using a portable test-strips reader designed for point-of-care measures in humans (Accutrend, Roche Diagnostics), whereas antioxidant capacity and ROM concentrations were measured through the OXY and d-ROMs tests, following the protocol described in the main text. The quantity of blood deposited on the test-strip was 10µL for triglycerides and glucose and 15µL for lactate. ROM concentration were not significantly correlated with triglycerides (Spearman's rank correlation test: $\rho$ = -0.01, $N$ = 76, $S$ = 74136, $P$ = 0.91), lactate ($\rho$ = 0.18, $N$ = 19, $S$ = 937, $P$ = 0.47), or glucose concentration ($\rho$ = -0.37, $N$ = 20, $S$ = 1817, $P$ = 0.11). There was no correlation between the total antioxidant capacity of the plasma and triglycerides ($\rho$ = -0.16, $N$ = 93, $S$ = 155677, $P$ = 0.12), lactate ($\rho$ = 0.30, $N$ = 21, $S$ = 1072, $P$ = 0.18) or glucose concentrations ($\rho$ = -0.09, $N$ = 24, $S$ = 2510, $P$ = 0.67) either.